# Evaluation of Attack Vectors and Risks in Automobiles and Road Infrastructure

John N. Brewer III
Department of Computer Science
& Information Technology
Hood College
Frederick, MD, USA
jnb3@hood.edu

George Dimitoglou
Department of Computer Science
& Information Technology
Hood College
Frederick, MD, USA
dimitoglou@hood.edu

*Abstract*— The evolution of smart automobiles and vehicles within the Internet of Things (IoT) - particularly as that evolution leads toward a proliferation of completely autonomous vehicles - has sparked considerable interest in the subject of vehicle/automotive security. While the attack surface is wide, there are patterns of exploitable vulnerabilities. In this study we reviewed, classified according to their attack surface and evaluated some of the common vehicle and infrastructure attack vectors identified in the literature. To remediate these attack vectors, specific technical recommendations have been provided as a way towards secure deployments of smart automobiles and transportation infrastructures.

## I. INTRODUCTION

In recent years there has been an exponential growth in complex electronic computer control systems for vehicles, which has been particularly evident in the evolution of 'smart cars' and even more so in the rise of 'autonomous vehicles' – i.e., those that can accelerate, steer, brake, and navigate without human input. These are vehicles of all sizes, from passenger cars and buses to trucks that have also become a growing area of interest within the context of the Internet of Things (IoT). Business forecasts predict a high degree of interest in IoT connected cars, creating the "Internet of vehicles" and a new transportation environment of driverless cars [1] [2].

Automobiles are a complex collection of mechanical and electrical systems and devices and have been so since their inception. In response to competition, changing customer needs and regulatory requirements, improvements to these systems have been continuous throughout the automobile's history. Many of the improvements included technologies invented specifically for the automobile, while others were originally created for other purposes but quickly found uses within the automotive realm. For example, the transistor and the junction type transistor were first demonstrated in 1947 and 1951 respectively. [3] By 1963, transistorized breaker-less ignition, using transistors instead of contact points for building/collapsing the magnetic coil field so as to induce sparks to the spark plugs, could be found as an option from two U.S. carmakers on select models, and available from at least 12 different aftermarket suppliers [4]. Even before that, in 1958, the integrated circuit (IC) was developed, containing multiple transistors [5]. By the early 1970's integrated circuits had become a staple of mainstream production automotive applications, controlling spark function in units such as the High Energy Ignition (HEI) modules from General Motors (GM). By the early 1980's computers had moved into the automotive arena with the Engine Control Unit (ECU) [6], a component interfacing with fuel and ignition systems along with a variety of sensors to receive feedback and make instantaneous engine control decisions such as air/fuel ratio and spark timing. ECU's are found now in every car, although today they use sophisticated computers to control steering, braking, climate control, navigation, infotainment, and much more [6].

The motivation behind this paper is to review landmark exploits in the area of automobile computer controls and supporting road infrastructure. Along with the physical security considerations, the new era of automobiles as part of the IoT, depends heavily on software, increasing the problem complexity and widening the possible attack vector beyond traditional, physical access boundaries. The dependence on software is significant, with the average modern high-end car containing over 100 million lines of code [6]. Through the exploration of known, past exploits, we hope to cover some interesting technologies, evaluate their use and identify their risks. This area of study becomes more complex given the realization that cars today are not only controlled by computers but are becoming networked not just internally with their own components, but also externally to other cars through vehicle-to-vehicle (V2V) networks and to the surrounding infrastructure of roadways, traffic control devices, signage, etc. in a vehicle-to-infrastructure (V2I) network environment. Considering the possibility of exploiting a security vulnerability in this landscape complicates the issues and affects everyone in the supply chain, from the manufacturer to the driver, along with non-autonomous vehicle drivers and pedestrians.

In this study we embark on providing a brief overview of the environment, present some of the most well-known attack vectors, and propose a set of technical recommendations for the secure and safe deployment of smart cars within smart infrastructures.

## II. THE ENVIRONMENT

In determining the attack vectors of autonomous vehicles, it is important to consider the inherent connectivity within the environment. Autonomous vehicles establish and maintain both internal and external connections. Internally, vehicles are connected with their own components such as a Tire Pressure Monitoring System (TPMS) or driver devices such as key fobs and mobile phones. Externally, vehicles could be connected to other vehicles forming V2V *ad hoc* networks, roadway infrastructure, forming vehicle-to-infrastructure V2I *ad hoc* networks and manufacturer's assistance platforms (e.g. OnStar). The types of connections are very diverse, ranging from Bluetooth, WiFi, near field communication (NFC), and global system of mobile communications (GSM) cellular protocols [7] to ultra-high frequency (UHF) radio band transmissions [8].

The high connectivity of autonomous vehicles, combined with the diversity of connected components make them an attractive target for vulnerability exploitation.

*A. Attack Vector: "Lock Picking"*

Keyless door entry allows users to open and lock car doors using a key fob with buttons that wirelessly transmit '*open*' and '*lock*' signals. The exploit is based on a *man-in-the-middle* attack using a device that captures the transmitted signal from the key fob and simultaneously sends a jamming signal on the same frequency. The key fob user, thinking their key fob malfunctioned, tries again; this time the car door unlocks. Key fobs transmit unique, one-time codes each time they are used. The attacker's device captured the first code, then transmits and "sacrifices" it to unlock the door as it captures the second code transmission. As the codes have no expiration time, the attacker may use the second captured code to transmit and open the door. The same technique and device may also be used to attack garage door openers [9].

These types of attacks have been successfully demonstrated on garage door openers and several cars including Chrysler, Daewoo, Fiat, GM, Honda, Toyota, Volvo, Volkswagen, and Jaguar [10]. The work behind these types of attacks can be explained mathematically [11] while several hardware devices exist to carry out these types of attacks. For example, the Rolljam device to attack remotely operated RF signals is small and fast, OpenSesame opens garage doors using a Mattel toy, OwnStar finds, unlocks, and starts GM OnStar cars.

The same technology has also been used in the amateur drone space with SkyJack, a Raspberry Pi drone that locates and attacks other drones [12]. These devices are widely available in the market, so much so that complaints were made about them to the UK's Home Office (the government's department charged with national security) but their sales continued unabated [13].

Protecting against these types of attacks could be addressed by the remote door control system manufacturers. Possible solutions could include the implementation of anti-jamming and code-grabbing resistant functions, along with updated chip firmware with time-expiring signals. The latter solution has been adopted by one manufacturer [14] but others have not yet followed suit.

*B. Attack Vector: Vehicle-Monitoring Components*

A well-studied exploitation has been related to the tire pressure monitoring system (TPMS) [8]. The TPMS transmits information to the vehicle's Electronic Control Unit (ECU), while continuously monitoring tire pressure to enable the ECU to trigger a dashboard warning light (in some vehicles the actual tire pressure is displayed) if tire pressure falls outside specifications. These systems are known to have been deployed without any security safeguards.

A possible passive attack can allow attackers to track vehicle movement and location using mobile or roadside tracking stations as TPMS signals can be captured up to 40 meters away from the vehicle. On the other hand, the lack of cryptographic protection of the TPS signals can enable an active attack as the vehicle's ECU trusts any information provided from vehicle sensors. It is therefore possible to wirelessly inject spoofed signals, tricking the ECU into displaying false and even out of range tire pressure measurements. Mitigating these types of attacks can be as simple as using hardware pairings (e.g. ECU communication with sensors of specified hardware addresses) and deploying encrypted-only signal transmissions.

*C. Attack Vector: Road. Infrastructure*

As vehicles become more connected, road infrastructure components become an intrinsic part of these growing transportation computing networks. In fact, vehicle-to-infrastructure (V2I) connectivity becomes an integral part of the environment which includes smart traffic lights, pedestrian crosswalk sensors and smart road, adaptive road signs.

Exploiting vulnerabilities of road signs to alter the displayed messages may sound more like an innocent prank but it can quickly become a very serious issue. Case in point, a 2014 attack when a foreign national initially compromised five overhead highway signs in North Carolina and six others in two other states, altering their displays by indicating they were hacked by SunHacker [15]. The attack was possible due to the negligence of the signal operators in changing the default password set by the sign manufacturer. [16]

Though individual road signs have been compromised and changed to warn of zombie attacks and other farcical threats, this is the first known case of a mass compromise of networked road signs. While the particular attack was benign, it is clear that compromised road infrastructure by a malicious actor with control over highway signage could cause major problems that may be more than traffic disruptions, particularly during an emergency.

Mitigating this type of attack requires following best practices in password management for the road signs as one would expect for any other computing devise that is shipped with a default password from the manufacturer. The demonstrated attacks provided misinformation to drivers but it is easy to extrapolate how misleading information could be passed to vehicles through their sensor payload from road infrastructure assets.

From the vehicle's perspective protection has to come from secure design of sensors that provide both V2V and V2I sensing related to speed, motion, and detection of road and traffic conditions. Each one of these sensors represents a potential attack vector, so secure design should be a consideration not only in deployment but also in sensor design. There are different approaches in sensor design to mitigate attacks. For example, developing smart sensors that use linear Gaussian dynamics techniques that are able to statistically profile expected sensor inputs and detect attacks [17].

*D. Attack Vector: Remote Attacks*

A widely known automotive attack was against a Jeep Cherokee [18]. The remote, internet-based attack was executed by exploiting a vulnerability found in the vehicle's entertainment system. The attacker was able to take control away from the driver and operate steering, brakes, and transmission controls as well as dashboard features and the vehicle's GPS system. The attack was based on exploiting Chrysler's smart phone application Uconnect. Once the car's IP address was captured, the vehicle became vulnerable via an Internet connection, requiring no physical connection or access. The root-cause of the vulnerability was based on the inherent weakness of the vehicle's wireless access point which was vulnerable to a password-guessing and brute force attack [19]. The attack prompted a 1.4 million vehicle recall in order to install a necessary software patch.

Another well-documented attack involves Nissan's electric car, the Leaf. Similar to the Jeep attack, the vulnerability was based on the vehicle's ability to be remotely accessed via a cell-phone using an application called NissanConnect EV (formerly known as CarWings) [20]. The vulnerability allowed an attacker to access not just one vehicle's battery status information and to control the heating and air conditioning system, but also to connect to other Leaf vehicles, around the world, using Vehicle Identification Numbers (VINs). Nissan responded to reports of this hack by quickly deactivating the application [21].

While these examples reveal serious remote access vulnerabilities, another car manufacturer, Tesla, seems to be less vulnerable to attacks, despite their significant dependence on software. While attackers have demonstrated an ability to connect to a Tesla and perform some functions, they had to overcome extensive security blocks, and they were able to succeed only through physical access and extensive physical tear-down to access necessary parts to compromise and gain control [22].

*E. Attack Vector: Vehicle Architecure Vulnerabilities*

A potential area for attacks in today's automobiles is the Controller Area Network (CAN) bus architecture used for the connection of vehicle internal components. Early automobiles used point-to-point wired connections until it became apparent that as car features and complexity grew (i.e., as cars became ECU-equipped) the amount of wiring required to continue connecting each component would become unwieldy and hard to maintain.

In 1985 an alternative architecture for automobile internal connectivity was developed, resembling the bus connectivity commonly used in personal computers [23]. The CAN bus architecture greatly reduced the amount of wiring for a vehicle and has been required to be deployed in automobiles since 2008 [24].

Modern automobiles tend to have two CAN buses, one providing low-speed and another high-speed communication. The low-speed bus controls and carries less critical systems and components such as heating and air conditioning, radio, theft deterrent modules and TPMS signals. The high-speed bus (the more trusted of the two in its network implementation) controls critical engine, powertrain, and braking system components.

These two networks are supposed to have strong connections and gateways between them. The gateways are only to be reprogrammable via the high-speed bus, ensuring a higher degree of protection that the less-trusted low-speed bus network. Attached to both of those buses are the body control and the telematics modules that are used for remote manufacturer access, such as GM's OnStar systems.

Within the CAN architecture, broadcast messages are sent on the network and each device "decides" if the message is relevant to it or not.

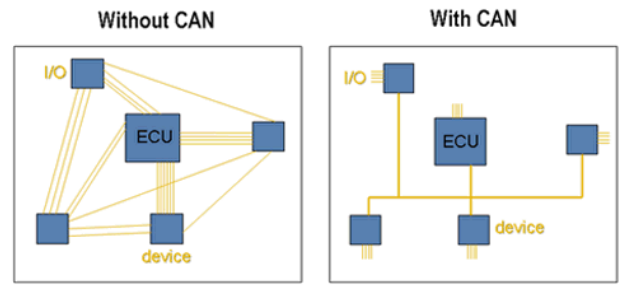

Fig. 1. *The introduction of the Controller Area Network bus architecture (right) significantly simplified traditional architectures (left) by eliminating the point-to-point connection of all communicating devices and the ECU (Image credit: National Instrument [23])*

However, while the technology is efficient in terms of reducing wiring complexity, it was designed to be lightweight, but not designed to segment the network and isolate components or to create boundary defenses. Given the centralized nature of the architecture, everything is connecting to the ECU [6]; once an attacker gains access to one part of the system, they are able to access and attack all other systems.

Another inherent weakness to the CAN architecture is the lack of device authorization, as there is no way to confirm that transmitted information originates from the actual device. In addition, the traffic within the CAN is unencrypted, thereby making message transmission visible once intercepted.

The best approach to remediate this inherent vulnerability is encrypting messages transmitted within the CAN and several such solutions exist such as SecureCAN which uses a three-pass algorithm of substitution, transposition, and time-multiplexing [6] and code obfuscation and encryption over the standardized On-Board Diagnostics II (OBD-II) interface with the ECU [25]. Device authentication can also be accomplished by designing device controllers that trust a specified list of known providers. Coupled with encrypted messaging the danger of a spoof attack is minimized.

One interesting study that was conducted in both laboratory and field settings, was to determine the security implications if there were to be a malicious attack on a vehicle's connected systems [26]. The result of this study was the identification of two primary vectors as prospective ways to launch a malicious attack. The first way is by accessing the physical On-Board Diagnostic II (OBD-II) port which is required for emissions testing and the second way is by accessing any available wireless interfaces. In the study, it was demonstrated how different exploits could be launched by connecting a laptop to the vehicle's OBD-II port. Exploiting CAN weaknesses, several systems and components were attacked. A particularly impressive attack included *reflashing* (i.e. reprogramming) the ECU's while driving, which should not be able to occur since ECUs are supposed to reject reflashing efforts whenever the engine is running. The result is that the car's engine stops running.

Several other attacks have been successfully demonstrated, such as programming the telematics unit via the low-speed CAN bus. This bus lacks authenticator fields, has weak access controls and message transmission is based on a broadcasting protocol, making it susceptible to Denial of Service (DOS) and other types of attacks [26].

Using the described inherent weaknesses, it was possible to demonstrate ECU *reflashing*, *fuzzing* (i.e. sending random unknown or partially known packets) to attack and confuse, study the system responses and be able to "reverse engineer" by observing how the system was handling various message.

In Table 1, there is a summary of some of the demonstrated exploits using this methodology and in particular CarShark, a custom-made CAN bus sniffer and packet injection software tool.

TABLE I.     SUMMARY OF EXPLOITS AGAINST THE CONTROLLER AREA NETWORK (CAN) BUS.

| Vulnerability | Attack Type | Result |
|---|---|---|
| Unauthorized access | Packet injection. Reflashing ECU while driving. | Engine stopped. Code loaded into vehicles' telematics unit. |
| Unauthorized access Denial of Service | Packet injection to Body Control Module (BCM). Fuzzing. | Door lock relay activated. Wipers turned on/forced off. Trunk opened. Horn activated. Auto-headlight control deactivated. Use of washer fluid. Brake/auxiliary lights rendered inoperable. |
| Unauthorized access | Packet injection to Engine Control Module (ECM) | Engine timing and RPM disturbed. Engine cylinders stopped. Grind starter motor |
| Unauthorized access | Packet injection to Brake Control Module (BCM) | Brake application and release (evenly or unevenly) at speeds below 5 mph. |
| Denial of Service | Packet injection to other CAN-connected bus devices. | Disable CAN bus communication. Freeze instrument panel status. |
| Unauthorized access | Packet injection | Kill engine |

Arguably, the CAN-based exploits may be unsettling although they were all the result of physical connections to the OBD-II link and required a full laptop computer within the victim's car to be able to seize control of the various automotive components. It would be reasonable to consider though that the physical laptop connection to the OBD-II could be replaced by a small form factor computer with wireless capability that could be easily concealed, opening the possibility for a remote network attack unbeknownst to the driver.

## III. FINDINGS AND RECOMMENDATIONS

The attack vectors described in the literature seem to apply equally to conventional smart cars and those featuring completely autonomous control. Both vehicle types share some type of an OBD diagnostic interface, similar CAN bus architecture, similar network connectivity using WiFi over known Internet protocols. They also already include or assume to be utilizing extensive sensory inputs from other vehicles and the surrounding road infrastructure. This connectivity increases the complexity of automotive systems, while also increasing their attack surfaces.

Despite the prevalence of smart vehicles and the numerous possible attack vectors identified and demonstrated in the literature for automobile systems in general, there is a distinct absence of real, malicious and successful attacks. Perhaps the reason is how expensive vehicles are, making them hard to be taken apart to better understand how they work and find potential vulnerabilities. Another explanation is the relative immaturity of roadway infrastructure to support large-scale adoption of smart vehicles and V2I connectivity and the even more limited communication capabilities across vehicles, rendering V2V an aspiration. This inherent immaturity and widespread adoption may be an opportunity for expert groups, lawmakers, manufacturers, engineers, software designers, and other involved stakeholders to better understand, design and implement technologies with security as a design feature. In general, the following recommendations could serve as a non-exhaustive account of remediations that would significantly reduce the inherent attack vectors in for automotive security:

- *Secure sensor design*. Secure sensor design to support autonomous vehicles are crucial to the function of these systems and present a current challenge and an opportunity. Innovative, cautious design that includes security mechanisms that preserve the authenticity of measurements by blocking injection attacks, ensure the identity of the source by preventing spoofing and safely transmit information (i.e. using encryption) within the vehicles and during communication with the environment will serve to reduce risks and attack vectors.
- *Secure control device design.* The cases of the code-grabbing devices that capture and exploit signals from key-fob transmission present a low-level risk that can be easily mitigated using both code expiration timestamps and source method authentication and pairing. To contextualize the sophistication and severity of this attack vector, they are equivalent to the use of a bent metal coat-hanger or the flattened "Slim Jim" tools widely sold at automotive supply stores 50 years ago. Both these legacy hand tools and today's code-grabbers perform the same task, gaining unauthorized car entry.
- *Authenticated, encrypted access to vehicle resources.* Both conventional and smart vehicles provide methods to perform deep (e.g. engine performance, timing) and shallow (e.g. tire pressure, cabin temperature) diagnostic checks either by physical or remote connection. In either case, access should be authenticated and the transmission of any information must be encrypted.
- *Authenticated, encrypted access to vehicle services.* Access to GPS, road assistance or even the vehicles'

entertainment system should be authenticated and the transmission of any information must be encrypted.

- **Secure road Infrastructure design.** Road infrastructure should follow best practices in the design of computer and network-based services, adhering to the principles of confidentiality, integrity and availability.
- **Paired, encrypted V2I communication.** Similar to cellular network communication that pairing and encrypted communication with cellular towers is based on a uniquely identifiable and traceable pairing of devices and antennas, vehicles should establish a similar type of connection with infrastructure assets.

## IV. CONCLUSION AND FUTURE WORK

Smart cars, particularly those of the imminently pending autonomous generation, hold the promise of making human lives easier and freeing our time for more productive endeavors. Moreover, they hold promise for dramatically reducing lost lives from traffic accidents resulting from the human proclivity for inattention and from our inherent lack of knowledge (i.e., inability to see far enough ahead to accurately predict pending collisions).

Computer controls using networked vehicles and infrastructures, far-reaching sensors, rapid and accurate decision making, and continuous attention to the task can virtually eliminate the risks inherent in human control of automobiles and other vehicles. While the prospect of human error is likely to remain as a continuous issue going forward, having a more clearly defined single set of best practices as they relate to the design and implementation of smart/autonomous vehicles and roadway infrastructure, along with a program of continuous training to help confirm compliance, will go a long way toward the mitigation of risk. The challenge will be to have in place a coherent set of standards and best practices and established well in advance of implementation, to minimize risk of hacking and the resultant damage to life and property from those who would do harm. Automotive cybersecurity is a relatively new topic and an emerging technology and will require much additional study

As with any technology, there are clearly risks associated with automotive and vehicle attacks, especially as cars become nodes of the IoT. The risk severity ranges from innocuous to potentially severe. A concerted effort between industry and government stakeholders to ensure the establishment and compliance to standard processes and best practices will do a lot towards reducing risks.

It has been very refreshing to see the automobile industry's approach to dealing with the security considerations of vehicles in this new era. Several automakers, in recognition of the complexity of the problem and to protect their brand and customers, have been very receptive about potential vulnerabilities. GM, Fiat/Chrysler and Tesla have gone as far as establishing bug bounty programs [27] [28] [29] that invite the reporting of vulnerabilities from the public while offering immunity from liability, monetary awards and public recognition, to those that discover vulnerability in their products. This is a very clever approach from the automakers, essentially recruiting a multitude of testers that can help improve vehicle security.

Real change, though, will rest in finding ways to ensure that both vehicles and the supporting infrastructure are thoughtfully designed to minimize risk before they become available for use.

In this paper we described a number of recommendations that may serve as inspiration for future work. It would be amiss to omit the need for policy development and legal requirements to support this new era of transportation. Securing vehicles and infrastructure will require extensive new or innovative designs in every aspect, which in turn, will require investments that manufactures may be reluctant to make unless they are required to do so. There are also issues of privacy given the amount of connected information that is being generated and distributed within this new generation of automotive networks and IoT. Therefore, it is necessary to examine and devise the legal rules, standards and frameworks that address all these issues that are beyond the scope in this study.